# CONCEPTUALISING NATURAL AND QUASI EXPERIMENTS IN PUBLIC HEALTH

Frank de Vocht[1,2,3], Srinivasa Vittal Katikireddi[4], Cheryl McQuire[1,2], Kate Tilling[1,5], Matthew Hickman[1], Peter Craig[4]

[1] Population Health Sciences, Bristol Medical School, University of Bristol, Bristol, UK

[2] NIHR School for Public Health Research, Newcastle, UK

[3] NIHR Applied Research Collaboration West, Bristol, UK

[5] MRC IEU, University of Bristol, Bristol, UK

[4] MRC/CSO Social and Public Health Sciences Unit, University of Glasgow, Bristol, UK

**Corresponding author:** Dr Frank de Vocht. Population Health Sciences, Bristol Medical School, University of Bristol. Canynge Hall, 39 Whatley Road, Bristol, United Kingdom. BS8 2PS. Email: frank.devocht@bristol.ac.uk. Tel: +44.(0)117.928.7239. Fax: N/A

**ABSTRACT**

**Background**

Natural or quasi experiments are appealing for public health research because they enable the evaluation of events or interventions that are difficult or impossible to manipulate experimentally, such as many policy and health system reforms. However, there remains ambiguity in the literature about their definition and how they differ from randomised controlled experiments and from other observational designs.

**Methods**

We conceptualise natural experiments in in the context of public health evaluations, align the study design to the Target Trial Framework, and provide recommendation for improvement of their design and reporting.

**Results**

Natural experiment studies combine features of experiments and non-experiments. They differ from RCTs in that exposure allocation is not controlled by researchers while they differ from other observational designs in that they evaluate the impact of event or exposure changes. As a result they are, in theory, less susceptible to bias than other observational study designs. Importantly, the strength of causal inferences relies on the plausibility that the exposure allocation can be considered 'as-if randomised'. The target trial framework provides a systematic basis for assessing the plausibility of such claims, and enables a structured method for assessing other design elements.

**Conclusions**

Natural experiment studies should be considered a distinct study design rather than a set of tools for analyses of non-randomised interventions. Alignment of natural experiments to the Target Trial framework will clarify the strength of evidence underpinning claims about the effectiveness of public health interventions.

# BACKGROUND

When estimating the causal effect of an intervention on some public health outcome the study design considered least susceptible to bias is the experiment (particularly the randomised controlled experiment (RCT)). Experiments have the distinctive feature that the researcher controls the assignment of the treatment or exposure. If properly conducted, random assignment balances known and unknown confounders between the intervention groups. In many evaluations of public health interventions, however, it is not possible to conduct randomised experiments and instead standard observational epidemiological study designs have traditionally been used which are known to be susceptible to unmeasured confounding.

Alternative designs that have become popular in public health, and which have distinct benefits over more traditional designs, are natural experimental designs, also known as quasi experiments or non-randomised intervention studies [1]. In these kinds of study, the allocation and dosage of treatment or exposure are not under the control of the researcher but are expected to be unrelated to other factors that cause the outcome of interest [2–5]. Such studies can provide strong causal information in complex situations, and examples are available in the literature showing they can generate effect sizes close to the causal estimates from RCTs. [6] [7] The term natural experiment is sometimes used synonymously with quasi-experiment, a much broader term that can also refer to researcher-led but non-randomised experiments. In this paper we argue for a clearer conceptualisation of natural experiments in public health research, and present a framework to improve their design and reporting.

Natural and quasi-experiments have a long history of use for evaluations of public health policy interventions. One of the earliest and best-known of which is the case of "Dr John Snow and the Broad Street pump"[8]. In this study, cholera deaths were significantly lower among residents served by the Lambeth water company, who moved their intake pipe from the Thames to a cleaner source, compared to those served by the Southwark and Vauxhall water company, who did not move their intake pipe. Since houses in that area were serviced by either company in an essentially random manner, this natural experiment provided strong evidence that cholera was transmitted through water [9].

**Natural and quasi experiments**

Natural and quasi experiments are appealing because they enable the evaluation of changes to a system that are difficult or impossible to manipulate experimentally; for example large events, pandemics and policy changes [6, 10]. They also allow for retrospective evaluation when the opportunity for a trial has passed [11]. Furthermore, these designs offer benefits over standard observational studies, which rely on the strong and often untenable assumption of no unmeasured confounding, because they exploit variation from a change in the exposure resulting from an exogenous event or intervention [13]. This aligns them to the '*do*-operator' in the work of Pearl[12]. Quasi and natural experiments thus combine features of experiments (exogenous exposure) and non-experiments (observations without a researcher-controlled intervention). As a result, quasi and natural experiments are generally less susceptible to bias than many other observational study designs.[13]

However, a common critique of quasi experiments is that because the processes producing variation in exposure are outside the control of the research team there is uncertainty as to whether biases in the estimation of the causal effects have been sufficiently minimized or avoided [6]. For example, a quasi experiment evaluation studied the impact of a voluntary change by a fast food chain to label its menus with information on calories on subsequent purchasing of calories. [14] Unmeasured differences in the populations that visit that particular chain compared to other fast food choices could lead to residual bias..

A distinction is sometimes made between quasi-experiments and natural experiments. The term 'natural experiment' has traditionally referred to the occurrence of an event with natural cause or a "force of nature" (Figure 1a) [1, 15]. These make for some of the most compelling studies of causation from non-randomised experiments. For example the Canterbury earthquakes in 2010-2011 could be used to study the impact of such disasters because about half of a well-studied birth cohort lived in the affected area with the remainder living outside [16]. More recently, the use of the term 'natural' has been understood more broadly as an event which did not involve the deliberate manipulation of exposure for research purposes, even if human agency was involved [17]. This differentiates it from other quasi experiments where, although an event is similarly studied, some aspect of the exposure allocation (but not full control) can be under the control of the researchers. A well-known example of a natural experiment thus defined is the "Dutch Hunger Winter" summarised by Lumey et al. [18], where the German authorities blocked all food supplies to the occupied West of the Netherlands resulting in widespread starvation. Food supplies were restored immediately after the country was liberated.. Because there was sufficient food in the occupied and liberated areas of the Netherlands before and after the Hunger Winter, exposure to famine occurred based on an individual's time and place (of birth) only. Similar examples of such 'political' natural experiment studies are the study of the impact of China's Great Famine [19] and the 'special period' in Cuba's history following the collapse of the Soviet Union and the imposition of a US blockade [20]. Natural experiments describing the study of an event which did not involve the deliberate manipulation of an exposure but involved human agency, such as the impact of a new policy, are the mainstay of 'natural experimental research' in public health, and the term natural experiment has become increasingly popular to indicate any quasi-experimental design (although it has not completely replaced it).

Dunning takes this concept further and defines a 'natural experiment' as a quasi-experiment where knowledge about the exposure allocation process provides a strong argument that allocation, although not deliberately manipulated by the researcher, is essentially random, referred to as 'as-if randomization' (Figure 1b) [4, 9] [7]. Under this definition, natural experiments differ from other quasi experiments, where the allocation of exposure, whether controlled by the researcher or not, does not clearly resemble a random process.

A third distinction between quasi and natural experiments has been made that argues that natural experiments describe unplanned events whereas quasi-experiments describe events that are planned (but not controlled by the researcher), such as policies or programmes specifically aimed at influencing an outcome (Figure 1c) [17].

When the assignment of units to exposures is not controlled by the researcher, with rare exceptions (for example lottery-system[21] or military draft[22] allocations), it is typically very difficult, to prove that true (as-if) randomization occurred. Because of the ambiguity of 'as-if randomisation' and the fact that the tools to assess this are the same as those used for assessment of internal validity in any observational study [11], the UK Medical Research Council (MRC) guidance advocates a broader conceptualisation of a natural experiment study as any study that investigates an event not under the control of the research team that divides a population into exposed and unexposed groups (Figure 1d).

Here, in agreement with all definitions but acknowledging the remaining ambiguity regarding the precise definition of a natural experiment [23], we argue that what distinguishes natural/quasi experiments from RCTs is that randomisation of allocation is not controlled by the researchers and what distinguishes them from other observational designs is that they specifically evaluate the impact of an event or change in exposure. The detailed assessment of the allocation mechanism (which determines exposure status) in this concept is essential and, therefore, demonstrating 'as if randomisation' will substantially strengthen any causal claims from natural experiment studies. The plausibility of this assumption thus relies on detailed knowledge of why and how units were assigned to conditions and how the assignment process was implemented, and this can be assessed quantitatively using standard tools for assessment of internal validity of a study [11], which are ideally supplemented by a qualitative description of the assignment process. Common with contemporary public health practice, we will use the term 'natural experiment', or NE, from hereon.

**METHODS**

Medline and Embase and Google Scholar were searched using search terms including quasi-experiment, natural experiment, policy evaluation and public health evaluation and key methodological papers were used to develop this work. Peer-reviewed papers were supplemented by grey literature.

**RESULTS**

**Part 1. Conceptualisations of Natural Experiments**

*A set of tools for non-randomised studies of exogenous exposures*

The MRC guidance places its emphasis on the analytic tools that are used to evaluate natural experiments. The implication of this conceptualisation is that NE studies are largely defined by the way in which they are analysed, rather than by their design. An array of different statistical methods is available to analyse NEs, including regression adjustments, propensity scores, difference-in-differences, interrupted time series, regression discontinuity, synthetic controls, and instrumental variables. Overviews including strengths and limitations of the different designs and analytic methods are provided in [11, 24].

*A Study Design*

The popularity of NEs has resulted in some conceptual stretching, where the label is applied to a research design that only implausibly meets the definitional features of a NE [9].(Spatial) observational studies exploring variation in exposures (rather than the study of an event) have sometimes also been badged as NE studies. A more stringent classification of NEs as a distinct study design rather than a collection of analytic tools is important because it prevents attempts to cover observational studies with a 'glow of experimental legitimacy' [9]. If the design rather than the statistical methodology defines a NE study, this opens up a plethora of statistical tools that can utilised to analyse the NE depending on knowledge about the event, its allocation process, and availability of data. Dunning argues that it is the research design, rather than the statistical modelling, that compels conviction when making causal claims and NEs can be divided into those where a plausible case for 'as-if random' assignment can be made and are thus able to adjust for unobservable confounding (which he defines as NEs), and those in which observable confounding is directly adjusted for through statistical means and their validity relies on the assumption that confounding is absent (and which he defines as 'other quasi experiments', and we define as 'weaker NEs') [7]; with the former considered more credible in theory [4]. In this framework, these 'as-if-randomised' NEs can be viewed as offering stronger causal evidence than other quasi-experiments because, in principle, they offer an opportunity for direct estimates of effects (akin to RCTs) where control for confounding factors would not necessarily be required [4], rather than relying on adjustment to derive conditional effect estimates [9]. Of course, the latter may well reach valid and compelling conclusions as well, but causal claims suffer to a higher degree from the familiar threats of bias and confounding.

Dawson and Sim further make a distinction between type 1 and type 2 NEs [23], and define a type 1 NE as a design in which the allocation of exposure is completely out of the control of the researcher, whereas in type 2 NEs researchers can have some control over intervention allocation; for example by being able to agree on a sequential rollout of an intervention or that the intervention is not initiated until a certain date. An example of such a 'type 2 study' is the evaluation of the 'Communities in Control of Alcohol' intervention in Greater Manchester, in which the research team were able to agree a sequential rollout of the intervention, but had little control over when it was implemented [25]. However, although these 'type 2' NEs have some intervention aspect (if not full control of the randomisation process) that differentiates them from other observational designs, most conceptualisations (with the exception of the MRC Guidelines, which do not make such a distinction) agree type 2 NEs are synonymous to quasi-experiments. An important aspect of this distinction in type 1 and type 2 NE studies (or natural and quasi experiment studies) is that whereas in NE studies (type 1) researchers have no control over the intervention, and so there are few ethical implications, in quasi experiments (type 2) studies researchers may have some influence, such as for example the delay of a new policy, which may have ethical implications that need to be considered[23].

**Part 2. A Target Trial Framework for Natural Experiment Studies**

In practice, there is considerable overlap between analytic and design-based uses of the term NE. If we consider NE studies a distinct study design however, they should be supplemented by a robust argument for 'as-if' random allocation of experimental conditions supported by both empirical evidence and by knowledge and reasoning about the causal question and substantive domain under question [7, 9]. Specifically, for claims of 'as-if' randomisation to be plausible, it must be demonstrated that the variables that determine treatment assignment are exogenous. This means that they are: i) strongly correlated with treatment status but are not caused by the outcome of interest (i.e. no reverse causality) and ii) independent of any other (observable and unobservable) causes of the outcome of interest [7]. Given this additional layer of justification, especially with respect to the qualitative knowledge of the assignment process and domain knowledge more broadly, we argue for co-produced evaluations of interventions in public health where feasible. If we appraise NEs as distinct study designs, which distinguish themselves from other designs because i) there is a particular change in exposure that is evaluated and ii) causal claims are supported by an argument of the plausibility of as-if randomization, then we guard against conflating NEs with other observational designs [26] [9].

There is a range of way of dealing with selection on observables and unobservables in NEs [7],[9] which can be understood in terms of the 'target trial' we are trying to emulate, had randomisation been possible [27]. The protocol of a Target Trial describes seven components common to RCTs [27]. In Table 1, we bring together elements of the Target Trial framework and conceptualisations of NEs[11] to derive an outline framework to describe the Target Trial for NEs. [11]

**DISCUSSION**

The Target Trial Approach for NEs outlined in Table 1 provides a relatively straightforward approach to identify NE studies that focusses on structural design elements and goes beyond the use of quantitative tools alone to assess internal validity[11]. It complements the ROBINS-I tool for assessing risk of bias in non-randomised studies of interventions, which similarly adopted the Target Trial framework [31].

An illustrative example of a well-developed NE study based on the criteria outlined in Table 1 is by Reeves et al. [32] in which the impact of the introduction of a National Minimum Wage on mental health was evaluated. The study compared a clearly defined intervention group of recipients of a wage increase up to 110% of pre-intervention wage, and compared these with clearly defined control groups of (1) people ineligible to the intervention because their wage at baseline was just above (100-110%) minimum wage and (2) people who were eligible, but whose companies did not comply and did not increase minimum wage. This study also included several sensitivity tests to strengthen causal arguments. We have aligned this study to the Target Trial framework in Additional file 1.

When there is not a plausible case for as-if randomisation and/or a suitable control group cannot be found, synthetic control methods can be used to support causal inference in non-randomised studies. Synthetic controls consist of weighted combinations of control units or predictor variables that serve as a counterfactual, to strengthen the causal claims in NE studies. For example, in a study aimed at evaluating the impact of alcohol licensing decisions on a variety of alcohol-related health and crime outcomes in the local area [33], synthetic controls were constructed from *a priori* defined comparable other local areas.

And finally, the framework does explicitly exclude observational studies that aim to investigate the effects of changes in behaviour without an externally forced driver to do so (the intervention), generally based on longitudinal datasets. For example, although a cohort study can be the basis for the evaluation of a NE study in principle, effects of the change of diet of some participants (compared to those who did not change their diet) is not an external cause (i.e. exogenous) and does not fall within the definition of an experiment [10]. However, such studies are likely to be more convincing than those which do not study within-person changes and we note that the statistical methods used may be similar to NE studies.

Nonetheless, NE studies remain based on observational data and thus biases in assignment of the intervention can never be completely excluded (although for plausibly 'as if randomised' NEs these should be minimal). It is therefore important that the quantitative degree of bias is reported, and the ROBINS-I risk of bias tool was specifically developed for this purpose [31]. It has additionally been argued that, because confidence intervals and statistical tests do not do this, sensitivity analyses are required to assess that no pattern of small biases could explain away any ostensible effect of the intervention [13]. Recommendations that would improve the confidence with which we can make causal claims from NEs have been outlined in Table 1, and to a large degree from work by Rosenbaum [13].

None of these recommendations outlined in Table 1 will by themselves eliminate potential bias in a NE study, but neither is it required to necessarily conduct all of these to be able to make a causal claim with some confidence (although if the data are available, it would be recommended nonetheless). Instead, a continuum of confidence in the causal claims that can be made based on the study design and the data is a more appropriate and practical approach [34]. Each sensitivity analysis aims to minimise ambiguity of a particular potential bias or biases, and as such a combination of selected sensitivity analyses can strengthen causal claims [13]. We would generally, but not strictly, consider RCTs as the design where we are most confident about such claims, followed by natural experiments, and then other observational studies; this would be an extension of the GRADE framework which currently only distinguishes between trials and observational studies [35].

Our recommendations are of particular importance for ensuring rigour in the context of (public) health research where natural experiments have become increasingly popular for a variety of reasons, including the availability of large routinely collected datasets. These invite the discovery of natural experiments, even where the data may not be particularly applicable to this design, but also these enable many of the sensitivity analyses to be conducted from within the same dataset or through linkage to other routine datasets.

Finally, alignment to the Target Trial Framework also links natural experiment studies directly to other measures of trial validity, including pre-registration, reporting checklists, and evaluation through risk-of-bias-tools [31]. This aligns with previous recommendations to use established reporting guidelines such as STROBE, TREND, [11] and TIDieR-PHP [36] for the reporting of natural experiment studies, and could be customized to specific research areas (for example, as developed here for a systematic review of quasi-experimental studies of prenatal alcohol use and birthweight and neurodevelopment [37]).

**CONCLUSIONS**

We aimed to conceptualise natural experiment studies as they apply to public health, and argue for the appreciation of natural experiments as a distinct study design rather than a set of tools for the analyses of non-randomised interventions. Although there will always remain some ambiguity about the strength of causal claims, there are clear benefits to harnessing natural experiment studies rather than relying purely on observational studies. This includes the fact that NEs can be based on routinely available data and that timely evidence of real-world relevance can be generated. The inclusion of a discussion of the plausibility of as-if randomisation of exposure allocation will provide further confidence in the strength of causal claims. Aligning natural experiments to the Target Trial framework will guard against conceptual stretching of these evaluations and ensure that the causal claims about whether public health interventions 'work' are based on evidence that is considered 'good enough' to inform public health action within the 'practice-based evidence' framework [38].

**List of Abbreviations**

RCT = Randomised Controlled Trial

NE = Natural Experiment

SUTVA = Stable Unit Treatment Value Assumption

ITT = Intention-To-Treat

**DECLARATIONS**

**Funding:** This study is funded by the National Institute for Health Research (NIHR) School for Public Health Research (Grant Reference Number PD-SPH-2015). The views expressed are those of the author(s) and not necessarily those of the NIHR or the Department of Health and Social Care. The funder had no input in the writing of the manuscript or decision to submit for publication. The NIHR School for Public Health Research is a partnership between the Universities of Sheffield; Bristol; Cambridge; Imperial; and University College London; The London School for Hygiene and Tropical Medicine (LSHTM); LiLaC – a collaboration between the Universities of Liverpool and Lancaster; and Fuse - The Centre for Translational

Research in Public Health a collaboration between Newcastle, Durham, Northumbria, Sunderland and Teesside Universities. FdV is partly funded by National Institute for Health Research Applied Research Collaboration West (NIHR ARC West) at University Hospitals Bristol NHS Foundation Trust. SVK and PC acknowledge funding from the Medical Research Council (MC_UU_12017/13) and Scottish Government Chief Scientist Office (SPHSU13 & SPHSU15). SVK acknowledges funding from a NRS Senior Clinical Fellowship (SCAF/15/02). KT works in the MRC Integrative Epidemiology Unit, which is supported by the Medical Research Council (MRC) and the University of Bristol [MC_UU_00011/3 ].

**Ethics approval and consent to participate:** Not applicable

**Consent for publication:** Not applicable

**Availability of data and materials:** Not applicable

**Competing interests:** The authors declare that they have no competing interests

**Authors' contributions**: FdV conceived of the study. All authors contributed to interpretation of theory and evidence. FdV wrote the first version of the manuscript. All authors provided input in subsequent versions and approved of the final manuscript.

**Acknowledgements**: Not Applicable

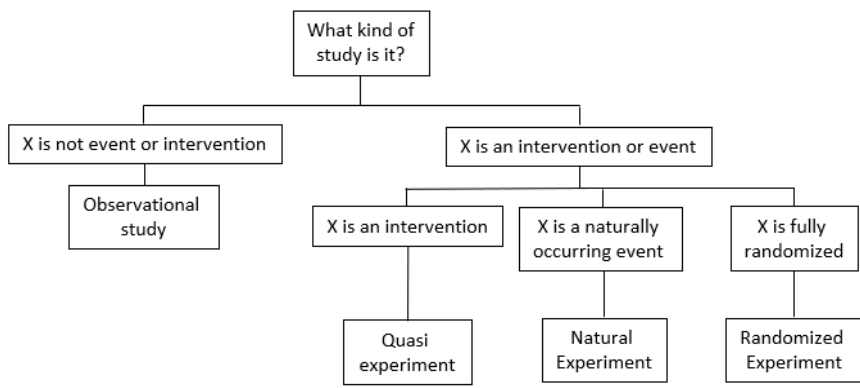

1a. Graphical overview of Shadish, Cook and Cambell[1]

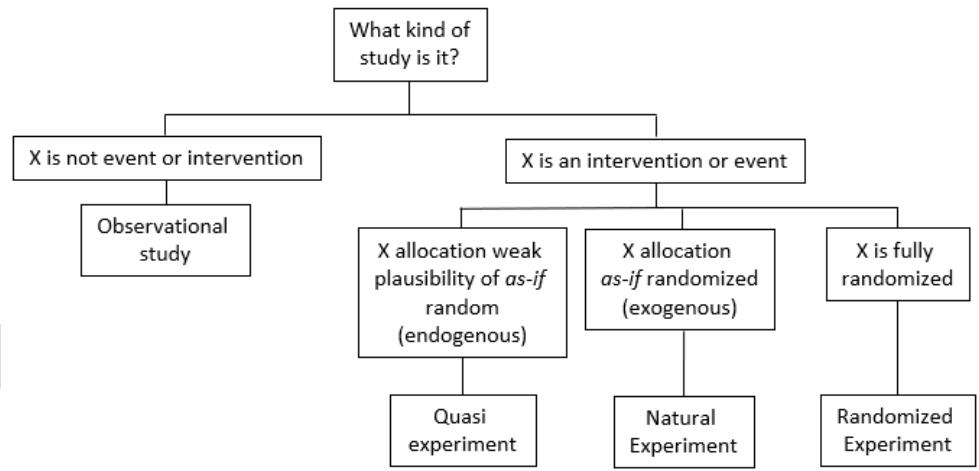

1b. Graphical overview of Dunning[4]

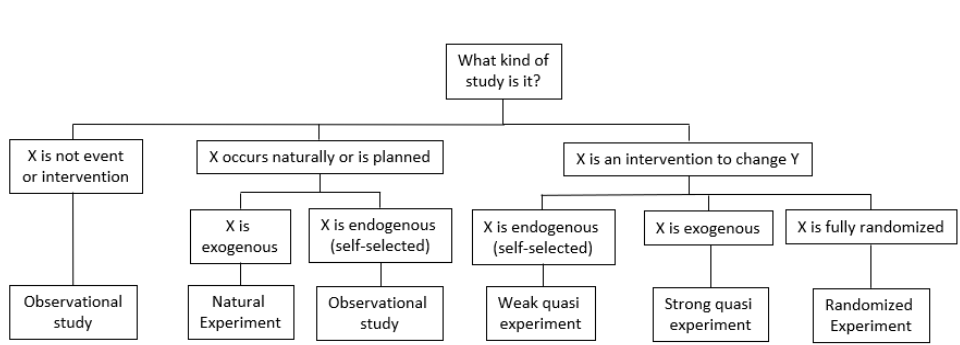

1c. Graphical overview of Remler and van Ryzin[17]

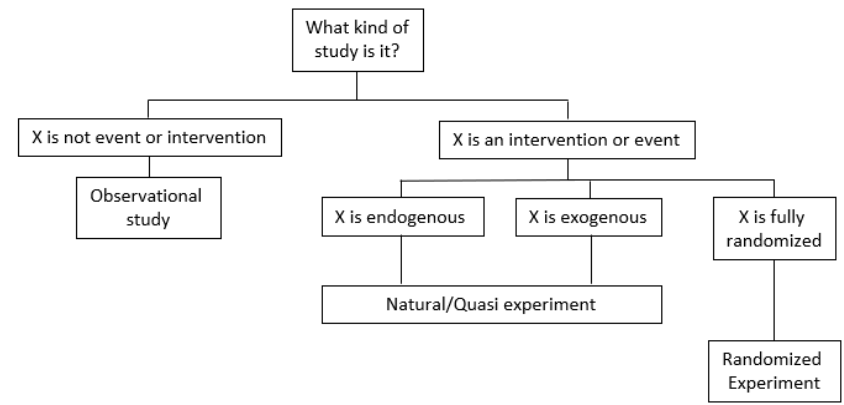

1d. Graphical overview of MRC Guidelines[5]

**Figure 1.** Different conceptualisations of natural and quasi experiments within wider evaluation frameworks.

*Table 1. Outline of the Target Trial for Natural Experiments*

| Protocol Component | Considerations for improvement of causal claims from Natural Experiments | Further recommendations for improvement of Natural Experiment Studies |
|---|---|---|
| Eligibility Criteria | <ul><li>A precise and detailed description of the population who have/will feasibly be exposed to the intervention, with special focus on the boundaries of the intervention which may be fuzzy and/or may not overlap with boundaries of (routine) data collection or risk of the outcome.</li><li>Define and describe eligibility of potential control populations to ensure independence and exclude spill-over effects. [28]</li></ul> | <ul><li>Consider broadening out the eligibility criteria for control groups to include, for example, comparable groups or areas from other geographical locations for sensitivity analyses.</li></ul> |
| Treatment strategies | <ul><li>Precisely define the intervention, the dose and treatment regimes, and what it aims to affect, including when and where it is introduced.</li><li>Define baseline timepoint, and consider pre-implementation changes resulting from anticipating the intervention (for example changes in behaviour or reactions from industry [29]). This requires pre-intervention data.</li><li>Define the control condition (including the potential for reactions even if intervention wasn't received) in the post-intervention period and/or precisely define the counterfactual</li><li>Describe plausibility of the Stable Unit Treatment Value Assumption (SUTVA) [30]</li></ul> | <ul><li>Consider other, likely earlier, baseline timepoint to exclude anticipation behaviour in sensitivity analyses</li></ul> |
| Assignment procedures | <ul><li>The assignment procedure of the intervention is not controlled by the researcher but *as-if* randomisation is plausible.</li><li>The plausibility of *as-if* randomization of unobserved factors in a natural experiment can only be assessed through detailed knowledge of the assignment rationale and procedures.</li><li>Intervention and control groups remain in their allocation group throughout the study</li><li>The intervention group can also be the whole population (e.g. if exposed to the intervention at a well-defined timepoint) or, in the absence of a suitable control population defined by a temporal or spatial boundary, can be a synthetic counterfactual</li><li>Conditional exchangeability can be formally evaluated for observed factors, but not unobserved factors. This requires knowledge about exposure allocation procedures.</li></ul> | <ul><li>Consider whether partial control of assignment of intervention is possible</li><li>Consider inclusion of control groups or synthetic counterfactuals to improve assessment of conditional exchangeability for observed and unobserved factors</li></ul> |
| Follow-up period | • Starts prior to assignment of intervention to groups, includes assignment, and ends after *a priori* defined period post-intervention | • Consider different follow-up periods to assess evidence of pulse impacts (short-term temporal effect followed by regression to the mean) |
| Outcome | • Study-specific *a priori* hypothesized individual-level or population-level parameter at *a priori* defined period post-intervention or cumulative/average outcome from start of intervention until *a priori* defined period post-intervention. | Consider different outcomes:<br>• also hypothesised to be affected by intervention (positive control)<br>• hypothesised to be unaffected by intervention (negative control) |
| Causal contrasts of interest | • Precisely define the causal contrast to be evaluated<br>• Natural experiments will enable estimation of intention-to-treat effects.<br>• Where possible, identify individuals who adhered to protocols and obtain per-protocol effects (although in natural experiments this information may be rarely available) | • Consider if per-protocol effects can be evaluated (possibly in subgroup) in addition to ITT<br>• Consider additional causal contrasts, for example in subgroups |
| Analysis plan | • Difference between post-intervention minus pre-intervention outcome of interest in intervention group and post-intervention minus pre-intervention outcome of interest in control. Can be expressed in a variety of relative and absolute measures and obtained using a variety of statistical methods. | • Temporal falsification analyses are conducted by choosing different, randomly assigned, implementation times for the intervention<br>• Spatial falsification analyses are conducted by repeated analyses with different combinations of units, irrespective of true assignments<br>• Analytic triangulation by analysing natural experiment using different statistical methods. |

# CONCEPTUALISING NATURAL AND QUASI EXPERIMENTS IN PUBLIC HEALTH

Frank de Vocht[1,2,3], Srinivasa Vittal Katikireddi[4], Cheryl McQuire[1,2], Kate Tilling[1,5], Matthew Hickman[1], Peter Craig[4]

[1] Population Health Sciences, Bristol Medical School, University of Bristol. [2] NIHR School for Public Health Research. [3] NIHR Applied Research Collaboration West. [5] MRC IEU. [4] MRC/CSO Social and Public Health Sciences Unit, University of Glasgow

# ONLINE SUPPLEMENTARY MATERIALS

This study is funded by the National Institute for Health Research (NIHR) School for Public Health Research (Grant Reference Number PD-SPH-2015). The views expressed are those of the author(s) and not necessarily those of the NIHR or the Department of Health and Social Care. The funder had no input in the writing of the manuscript or decision to submit for publication. The NIHR School for Public Health Research is a partnership between the Universities of Sheffield; Bristol; Cambridge; Imperial; and University College London; The London School for Hygiene and Tropical Medicine (LSHTM); LiLaC – a collaboration between the Universities of Liverpool and Lancaster; and Fuse - The Centre for Translational Research in Public Health a collaboration between Newcastle, Durham, Northumbria, Sunderland and Teesside Universities. FdV is partly funded by National Institute for Health Research Applied Research Collaboration West (NIHR ARC West) at University Hospitals Bristol NHS Foundation Trust. SVK and PC acknowledge funding from the Medical Research Council (MC_UU_12017/13) and Scottish Government Chief Scientist Office (SPHSU13 & SPHSU15). SVK acknowledges funding from a NRS Senior Clinical Fellowship (SCAF/15/02). KT works in the MRC Integrative Epidemiology Unit, which is supported by the Medical Research Council (MRC) and the University of Bristol [MC_UU_00011/3 ].

*Table 1. the Target Trial for Natural Experiments and Reeves et al.*[32]

| Protocol Component | Considerations for improvement of causal claims from Natural Experiments | | Further recommendations for improvement of Natural Experiment study, based on the Target Trial Framework | |
|---|---|---|---|---|
| | **Protocol criterion** | **Reeves et al. evidence towards meeting criterion** | **Protocol recommendation** | **Reeves et al. evidence towards meeting recommendation** |
| Eligibility Criteria | A precise and detailed description of the population who have/will feasibly be exposed to the intervention, with special focus on the boundaries of the intervention which may be fuzzy and/or may not overlap with boundaries of (routine) data collection and risk of the outcome. | The study population are subjects enrolled in the British Household Panel Survey. A nationally representative longitudinal survey of 5,500 households and ~10,000 individuals (More detail [39]), and includes men and women aged 22-59 who worked at least 1hr per week in 1998 and 1999. | | |
| | Define and describe eligibility of potential control populations to ensure independence and exclude spill-over effects | The intervention group were individuals who received the minimum wage increase and the control population were those who did not, either because they were ineligible or because their employer did not comply with the legislation.<br><br>Introducing a minimum wage may also increase wages for those who are just above the minimum wage threshold, and so, those who are untreated could also be influenced by the intervention. While such spill-over effects are theoretically plausible, and this appears to have occurred | Consider broadening out the eligibility criteria for control groups to include, for example, comparable groups or areas from other geographical locations for sensitivity analyses. | |
| | | in the USA, previous studies suggest that they did not influence the UK's wage distribution. | | |
| Treatment strategies | Precisely define the intervention, the dose and treatment regimes, and what it aims to affect, including when and where it is introduced. | The intervention was the introduction of the UK National Minimum Wage brought into force on April 1st, 1999. This was aimed to affect all workers who earned less than £3.60 per hour in 1998.<br>The intervention group comprises those who earned less than £3.60 per hour in 1998 and who then earned between £3.60 and £4.00 per hour in 1999. | Consider other, likely earlier, baseline timepoint to exclude anticipation behaviour in sensitivity analyses | Robustness analyses were conducted using standard regression methods using an earlier start point of 1994.<br><br>For reference: the policy was a key policy for Labour in the 1997 election and a key piece of legislation in 1998 |
| | Define baseline timepoint, and consider pre-implementation changes resulting from anticipating the intervention (for example changes in behaviour or reactions from industry [29]). This requires pre-intervention data. | The baseline timepoint was wave 8 of the data collection, which started in September 1998. | | |
| | Define the control condition including potential reactions in the post-intervention period and/or precisely define the counterfactual | The control condition is people ineligible to the intervention because their wage at baseline was just above (100-110%, or (£3.60 to £4 per hour) the threshold to receive Minimum Wage.<br><br>A second control group was used in the study, and were people who were eligible, but whose companies did not comply with the National Minimum Wage legislation and did not increase minimum wage. | | |
| | Describe plausibility of the Stable Unit Treatment Value Assumption (SUTVA) [30] | | | |
| Assignment procedures | The assignment procedure of the intervention is not controlled by the researcher but *as-if* randomisation is plausible. | The primary control group were people ineligible to the intervention because their wage at baseline was just above (100-110%) the threshold to receive Minimum Wage. | Consider whether partial control of assignment of intervention is possible | |
| | The plausibility of a*s-if* randomization of unobserved factors in a natural experiment can only be assessed through detailed knowledge of the assignment rationale and procedures | Selection in either group based on the arbitrary threshold is plausibly independent from other factors and can be considered exogenous, or as-if random (an instrumental variable (IV)). The plausibility is less strong for control group 2 because allocation relies on compliance of companies | | |
| | | | Consider inclusion of (additional) control groups or synthetic counterfactuals to improve assessment of conditional exchangeability for observed and unobserved factors | A second control group was used in the study, and were people who were eligible, but whose companies did not comply with the National Minimum Wage and did not increase minimum wage. Both control groups were combined in some analyses |
| | Intervention and control groups remain in their allocation group throughout the study | While in theory it is possible that persons above the income threshold could choose to move into a lower income group, this was judged very unlikely. | | |
|  |  | There was zero attrition from 1998 to 1999. |  |  |
|  | The intervention group can also be the whole population (e.g. if exposed to the intervention at a well-defined timepoint) or, in the absence of a suitable control population defined by a temporal or spatial boundary, can be a synthetic counterfactual |  |  |  |
|  | Conditional exchangeability can be formally evaluated for observed factors, but not unobserved factors. This requires knowledge about treatment allocation procedures. | To verify this assumption statistical tests to examine this as-if random assignment procedure were conducted [for observables]. |  |  |
| Follow-up period | Starts prior to assignment of intervention to groups, includes assignment, and ends after *a priori* defined period post-intervention | Follow-up starts at wave 8 of the data collection (1998), and post-intervention data are for wave 9 (start September 1999). | Consider different follow-up periods to assess evidence of pulse impacts (short-term temporal effect followed by regression to the mean) | Assessed whether positive mental health effects were sustained by adding a subsequent year, 2000/2001. |
| Outcome | Study-specific *a priori* hypothesized individual-level or population-level parameter at *a priori* defined period post-intervention or cumulative/average outcome from start of intervention until *a priori* defined period post-intervention. | The primary outcome was the probability of having a mental health problem assessed using the General Health Questionnaire 12 item version (GHQ-12), analysed as a continuous variable. | Consider different outcomes also hypothesised to be affected by intervention (positive control) | Additional outcomes were (from GHQ-12):<br>• 'constantly under strain'<br>• 'unhappy or depressed'<br>• self-reported depression<br><br>Additional outcomes also sensitive to short-term fluctuations in outcomes:<br>• self-reported elevated blood pressure<br>• number of cigarettes smoker per day among current smokers<br><br>Analyses were repeated with different calculations of the outcome:<br>• different overtime premiums<br>• a wage gap estimator |

|  |  |  |  | Additional outcomes hypothesised to be unaffected by intervention were (from GHQ-12)<br>• Self-reported chronic conditions (e.g. hearing difficulties) |
|---|---|---|---|---|
|  |  | Consider different outcomes also hypothesised to be unaffected by intervention (negative control) |  |  |
| Causal contrasts of interest | Precisely define the causal contrast to be evaluated | Change in GHQ-12 total scores from 1998 to 1999, specific components of the GHQ-12 (feelings of being 'constantly under strain' or 'unhappy or depressed' and a BHPS measure of self-reported depression between the intervention and control group. | Consider other causal contrasts, for example in subgroups |  |
|  | Natural experiments will enable estimation of intention-to-treat effects.<br><br>Where possible, identify individuals who adhered to protocols and obtain per-protocol effects (although in natural experiments this information may be rarely available) | The study was considered a per-protocol design. If someone earned below the threshold they would receive the new Minimum Wage (in the intervention group), unless their employer did not comply, while higher increases would indicate wage increases through other means | Consider if per-protocol effects can be evaluated (possibly in subgroup) in addition to ITT |  |
| Analysis plan | Difference between post-intervention minus pre-intervention outcome of interest in intervention group and post-intervention minus pre-intervention outcome of interest in control. Can be expressed in a variety of relative and absolute measures. | Difference-in-difference modelling framework.<br><br>Both differenced models and fixed-effects regression models including an interaction term between a period dummy and an intervention indicator.<br><br>Models were adjusted for age, sex, social class, and education. | Temporal falsification analyses are conducted by choosing different, randomly assigned, implementation times for the intervention | Estimated a series of random-effects linear regression models (controlling for time dummies) that include observations from 1994 to 2001 |
| | | | Spatial falsification analyses are conducted by repeated analyses with different combinations of units, irrespective of true assignments | |
| | | | Analytic triangulation by analysing natural experiment using different statistical methods. | Analyses were repeated with the natural log of the dependent variable<br><br>Natural experiment results were compared with traditional multivariable regression models. |